\documentclass[a4paper, 10pt, conference]{IEEEtran}  

\IEEEoverridecommandlockouts                              
\usepackage{listings}
\usepackage{graphicx}
\usepackage{hyperref}

\title{\LARGE \bf
Implementing distributed $\lambda$-calculus interpreter
}

\author{\IEEEauthorblockN{Alexandr Basov, 
Daniel de Carvalho, Manuel Mazzara}
\IEEEauthorblockA{
Innopolis University, Russia\\
Email: a.basov@innopolis.ru, d.carvalho@innopolis.ru, m.mazzara@innopolis.ru}}

\begin{document}

\maketitle
\thispagestyle{empty}
\pagestyle{empty}

\begin{abstract}

This paper describes how one can implement distributed $\lambda$-calculus interpreter from scratch. At first, we describe how to implement a monadic parser, than the Krivine Machine is introduced for the interpretation part and as for distribution, the actor model is used. In this work we are not providing general solution for parallelism, but we consider particular patterns, which always can be parallelized. As a result, the basic extensible implementation of call-by-name distributed machine is introduced and prototype is presented. We achieved computation speed improvement in some cases, but efficient distributed version is not achieved, problems are discussed in evaluation section. This work provides a foundation for further research, completing the implementation it is possible to add concurrency for non-determinism, improve the interpreter using call-by-need semantic or study optimal auto parallelization to generalize what could be done efficiently in parallel.

\end{abstract}

\section{Introduction}
\label{chap:intro}

The goal of this work is to implement a distributed $\lambda$-calculus interpreter. $\lambda$-calculus is a computational model introduced by the mathematician Alonzo Church in the 1930s as part of an investigation into the foundations of mathematics. This model is powerful enough to simulate any single-taped Turing machine. $\lambda$-calculus has applications in many different areas in mathematics, philosophy \cite{sep-type-theory}, linguistics \cite{bunt2008computing}, and computer science \cite{mitchell2003concepts}. $\lambda$-calculus has played an important role in the development of the theory of programming languages. Functional programming languages implement the $\lambda$-calculus.

To be more specific there is Entscheidungsproblem \cite{hilbert1950principles}, the problem asks for an algorithm that could take a statement of first-order logic as an input and tell whether it is true or false. But in order to answer this question there should be formalized a concept of algorithms that was done later by Alonzo Church with $\lambda$-calculus, Alan Turing with Turing Machine \cite{turing1936computable}. They proved that there is no general solution for this problem. Also these computational models were used to prove solutions for other problems, such as Halting problem.

Later it was shown that $\lambda$-calculus is logically inconsistent \cite{kleene1935inconsistency}, that is why Alonzo Church proposed simply typed $\lambda$-calculus, the system that has types in its notation. This system looks like intuitionistic logic, this similarity was generalized by Curry-Howard correspondence \cite{hindley1980hb} - it is the observation that two families of seemingly unrelated formalisms - namely, the proof systems (formal logic) on one hand, and the models of computation on the other — are in fact the same kind of mathematical objects. The simply typed $\lambda$-calculus corresponds to intuitionistic natural deduction.

This correspondence is used in functional languages and proving systems, Haskell and Coq, for example. Research about computation time of $\lambda$-terms \cite{phddeCarvalho, deCarvalho2017execution} is nowadays very active, which could be applied to prove complexity of algorithms.

One of the features that provides $\lambda$-calculus is referential transparency \cite{sondergaard1990referential} - this term describes the behavior when expression could be replaced with corresponding value without changing program behavior. And every expression in $\lambda$-calculus is referentially transparent, so using this feature it becomes easy to implement distributed computation for $\lambda$-calculus. 

\section{Objectives}
As a base for this project a $\lambda$-calculus interpreter represented by Krivine's machine \cite{krivine2007call} was taken. It is one of the abstract machines that could be used to implement functional languages. Then this interpreter will be extended to do distributed computations.

As the project intends to be a proof of a concept, it should have command-line interface with its own syntax. Thus the syntax must be designed and interpreter implemented according to that syntax. In this work we use Krivine's notation and "$\backslash$" symbol to replace $\lambda$.

$\lambda$-calculus brings a feature to computation where order of computation does not matter. Thus that feature could be used to distribute computation. In the implementation will be used the actor model \cite{hewitt1973session}.

\section{Design}
\label{chap:met}
In this section will be discussed the basis used during implementation. Considering that $\lambda$-calculus is a language, it could be splitted to implementation of parser and interpreter. After that parallelism and distribution is added. For each task there is a corresponding section. Also there is a section which discusses language decision for implementation.

\subsection{Language}
Generally speaking most of programming languages could be used for implementation. For this work Haskell language was chosen for $\lambda$-calculus interpreter implementation. There are several features that Haskell provides that we found useful.

Firstly, we found Algebraic data types useful. Algebraic data type could be describe as a combination of product types and sum types, also called tagged unions. Firstly we should introduce unit type:
\begin{lstlisting}[language=Haskell]
data () = ()
\end{lstlisting}
On the left side we have type declaration, this type has a name \textbf{()} and on the right side, we list type's constructors which starts with a name and could be followed with arguments. This type has no polymorphic arguments, and could be seen as a type where only 1 value is possible - \textbf{()}. We can use it as a neutral element for multiplication (\textbf{$a * 1 = a$}). Bellow is an illustration of the pair type:
\begin{lstlisting}[language=Haskell]
data Pair a b = Pair a b
\end{lstlisting}
This type has two polymorphic arguments, if we assume that \textbf{a} and \textbf{b} are sets of values of some type \textbf{a} or \textbf{b}. Thus \textbf{Pair} contains values of Cartesian product of those sets. \textbf{Pair} is used as multiplication in algebraic data types, and we can multiply two types:
\begin{lstlisting}[language=Haskell]
Pair Int Bool
\end{lstlisting}
And for this type all possible pairs of \textbf{Int} and \textbf{Bool} would satisfy the constraint. If we consider previously defined type \textbf{()}, we can arrive to the equality:
\begin{lstlisting}[language=Haskell]
Pair Int () = Int
\end{lstlisting}
This multiplication of \textbf{Int} and \textbf{()} is equal to \textbf{Int}, but up to isomorphism, by mentioning isomorphism we mean that there are two functions from \textbf{Int} to \textbf{Pair Int ()} and vice versa, and therefore we can provide these functions:
\begin{lstlisting}[language=Haskell]
toPair :: Int -> Pair Int ()
toPair n = Pair n ()

fromPair :: Pair Int () -> Int
fromPair (Pair n _) = n
\end{lstlisting}
In the \textbf{fromPair} above function pattern-matching is used. It is one of the features that we found valuable from using Haskell. In type definition we provide a structure and using this structure we can de-structure the argument to extract the value we are interested in.

In algebra we have addition. In terms of algebraic data types it is a sum type or as it was mentioned - tagged union. In set theory by union we mean set \textbf{C} where elements are elements of set \textbf{A} or \textbf{B}. But by tagged union we mean that we form a set of elements where each element is tagged with a set where this element came from. Consider the type:
\begin{lstlisting}[language=Haskell]
data Either a b = Left a | Right b
\end{lstlisting}
Operator "\textbf{$|$}" exactly follows this semantic. Values of type \textbf{Either} is a union of values of type \textbf{a} and \textbf{b}, but values are tagged with \textbf{Left} for values of type \textbf{a} and \textbf{Right} for values of type \textbf{b}. For addition we also should have the neutral element ($a + 0 = a$). For algebraic data types it is type \textbf{Void}:
\begin{lstlisting}[language=Haskell]
data Void
\end{lstlisting}
This type is only declared but it has no constructors, so it has no values. We can now move to addition, considering:
\begin{lstlisting}[language=Haskell]
Either Int Void = Int
\end{lstlisting}
Similar to multiplication this statement is true up to isomorphism. Now we should provide functions for isomorphism:
\begin{lstlisting}[language=Haskell]
fromInt :: Int -> Either Int Void
fromInt n = Left n

toInt :: Either Int Void -> Int
toInt (Left n) = n
\end{lstlisting}
But there are no functions for \textbf{Void} because we has no values in the \textbf{Void} type, therefore we can provide function \textbf{absurde} that takes \textbf{Void} and returns any value of any type where the trick is to pass value of type \textbf{Void} to the function, although it is not possible.

This possibility of summing-up is useful for defining possible expressions in language, in some sense we can express Backus–Naur form in algebraic data types.

The next part is exponentiation, we can raise a value to the power of a natural number, for example $a^2=a * a$. Now let us move to the list type, list can be seen as all possible tuples of all possible lengths $List a = () | (a , a) | (a, a, a) ...$ And to express this we can reference the \textbf{List} itself:
\begin{lstlisting}[language=Haskell]
data List a = Nil | Cons a (List a)
\end{lstlisting}
\textbf{Nil} is a constructor that equals to \textbf{()} up to isomorphism, and in the alternative case we have a recursive part, and it could be seen as $List(a) = () | a^1 | a^2 | a^3 ...$ It is also possible to raise a type \textbf{a} to the power of a type \textbf{b} $a^b$. This just means a function from type \textbf{a} to type \textbf{b}. With respect to sum types this feature is also helping to introduce possible expressions in a language.

In the beginning of the section it was mentioned that it is possible to use many languages. Statically typed languages help a lot when changes are done. Pattern matching and algebraic data types help to express a developed language, which expressions are possible and how the program should handle them. These properties have a lot of languages too, but we have chosen Haskell.

\subsection{Parser}
For parser implementation we used, following \cite{Wadler:1995:Monads}, recursive descent parsing approach. This kind of parsers could be expressed using monadic parsing \cite{hutton1996monadic}. The idea is that parser is a function, which takes string on input and returns a list of results, empty list of results denotes failure of a parser. This is the type:
\begin{lstlisting}[language=Haskell]
data Parser a = 
  Parser (String -> [(a, String)])

parse :: Parser p -> p
parse (Parser p) = p
\end{lstlisting}
When string is parsed, there is a list of results where each result is a tuple of a parsed value of type \textbf{a} and a prefix string which is not parsed yet. Returning a list allows us to build
parsers for ambiguous grammar, with many results being returned if the argument
string can be parsed in many different ways. And the parse function is used to return parser function from the Parser type.
The monadic part came from \textbf{Monad} type class:
\begin{lstlisting}
class Monad m where 
    return :: a -> m a
    (>>=)  :: m a -> (a -> m b) -> m b
\end{lstlisting}
This definition says that in order for a type to be a monad there should be implemented functions \textbf{return} and \textbf{$>>=$} for this type. For the \textbf{Parser}, implementation is straightforward:
\begin{lstlisting}[language=Haskell]
($) :: (a -> b) -> a -> b
f $ g = f g

instance Monad Parser where
    return a = Parser $ \cs -> [(a,cs)]
    p >>= f  = Parser $ \cs -> 
        concat [parse (f a) cs' |
               (a,cs') <- parse p cs]
\end{lstlisting}
\textbf{p} is a parser with parameter \textbf{a} and function \textbf{f} takes value of type \textbf{a} and produces new parser with parameter of type \textbf{b}. As a result, we should receive new parser with type \textbf{b} as a parameter. Thus implementation is tarted by defining new \textbf{Parser} which accepts string \textbf{cs}. This string passed to parser \textbf{p} which produces a list of \textbf{(a,cs')}, then we apply function \textbf{f} to each \textbf{a} and run parser with string \textbf{cs'}. In the end we receive a list of lists of results' tuples, then we eliminate outer list by using function \textbf{concat} on inner lists.

This is used to combine parsers and to have natural operational reading. For example, consider that we have a function \textbf{item :: Parser Char} where parser takes a string and takes the first character as a result. It could be used to build a parser that takes the first two characters and returns a tuple of two characters as a result:
\begin{lstlisting}
twoChars :: Parser (Char, Char)
twoChars = 
    item >>= \ch1 -> 
    item >>= \ch2 ->
    return (ch1, ch2)
\end{lstlisting}
We use a parser \textbf{item} that produces result \textbf{ch1}, then use the parser \textbf{item} again to produce \textbf{ch2} and return a tuple as a result. As we can see, in the code above parsing goes in natural order.

The next step is to define opportunity to combine different parsers so if the first parser fails, the second parser returns result. This could be achieved using \textbf{Monoid} type class.
\begin{lstlisting}[language=Haskell]
class Monoid a where
    mempty  :: a
    mappend :: a -> a -> a
\end{lstlisting}
Implementation is straightforward:
\begin{lstlisting}[language=Haskell]
instance Monoid Parser where
    mempty = Parser (\cs -> [])
    mappend p q = Parser $ \cs -> 
        parse p cs ++ parse q cs
\end{lstlisting}
Where \textbf{++} is a concatenation function for the \textbf{List} type. As we are interested only in first successful result, we define function \textbf{+++}:
\begin{lstlisting}[language=Haskell]
(+++) :: Parser a -> Parser a -> Parser a
p +++ q = Paresre $ \cs -> 
    case parse (mappend p q) cs of
        [] -> []
        (x:xs) -> [x]
\end{lstlisting}

The next important part is to have an validity check if a parsed character satisfies certain condition:
\begin{lstlisting}
sat :: (Char -> Bool) -> Parser Char
sat p = item >>= \c -> 
    if p c 
    then return c 
    else return mempty
\end{lstlisting}
This function could be used to parse digits, upper/lower-case letters. For example, this is a parser of particular character:
\begin{lstlisting}
char :: Char -> Parser Char
char c = sat (c ==)
\end{lstlisting}
From that point, it is possible to define more complicated parsers using recursion and defined parsers.

\subsection{Interpreter}
In order to discuss interpreter, firstly a language should be defined. We are using Krivine's notation for $\lambda$-terms. The language could be described as the set of syntax rules (the syntax is chosen according to Krivine's notation \cite{krivine1993lambda, krivine2007call}), table \ref{terms-rules} and the set of reduction operations, table \ref{reduction-operations}.

\begin{table}[h!]
\centering
\caption{Rules for defining terms in lambda-calculus}
\label{terms-rules}
\begin{tabular}{|l|l|p{5.2cm}|}
\hline
\textbf{Syntax} & \textbf{Name} & \textbf{Description} \\ \hline
x & Variable & A character or string representing a value \\ \hline
$\lambda$xt & Abstraction & Function definition (t is a $\lambda$-term). The variable x becomes bound in the expression. \\ \hline
(t)u & Application & Applying a function to an argument. t and u are $\lambda$-terms. \\ \hline
\end{tabular}
\end{table}

\begin{table}[h!]
\centering
\caption{Reduction operations for lambda-calculus}
\label{reduction-operations} 
\begin{tabular}{|l|l|p{3.8cm}|}
\hline
\textbf{Syntax} & \textbf{Name} & \textbf{Description} \\ \hline
$\lambda$xt{[}x{]} $\rightarrow$ $\lambda$yt{[}y{]} & $\alpha$-conversion & Renaming the bound variables in the expression. Used to avoid name collisions. \\ \hline
($\lambda$xt)u $\rightarrow$ t{[}x:=u{]} & $\beta$-reduction & Substituting the bound variable by the argument expression in the body of the abstraction \\ \hline
\end{tabular}
\end{table}

On top of syntax and reduction rules, there is a definition of free variables, defined inductively as:
\begin{itemize}
\item The free variables of $x$ are just $x$
\item The set of free variables of $\lambda$xt is the set of free variables of $t$, but with $x$ removed
\item The set of free variables of $ts$ is the union of the set of free variables of $t$ and the set of free variables of $s$.
\end{itemize}
Other variables are bounded variables.

The idea of interpretation is simple, for a given $\lambda$-term we compute a normal form and it could be defined in different ways and in this work we are using $\beta$-normal form. The definition of $\beta$-normal form is: if for a $\lambda$-term we can not apply $\beta$-reduction this term is in the $\beta$-normal form. We call a redex a $\lambda$-term for which we can apply $\beta$-reduction.

There are several evaluation strategies that could be used. The distinction between reduction strategies relates to the distinction in functional programming languages between eager evaluation and lazy evaluation:
\begin{itemize}
\item Full $\beta$-reduction. Any redex can be reduced at any time. This means essentially the lack of any particular reduction strategy.

\item Applicative order. The rightmost, innermost redex is always reduced first. Intuitively this means that function's arguments are always reduced before the function itself. Applicative order always attempts to apply functions to normal forms, even when this is not possible.

\item Call-by-name. The leftmost, outermost redex is always reduced first. That is, whenever possible the arguments are substituted into the body of an abstraction before the arguments are reduced.

\item Call-by-value. Only the outermost redexes are reduced: a redex is reduced only when its right hand side has been reduced to a value (variable or $\lambda$-abstraction).
\end{itemize}

In this work we are implementing the Krivine Machine, which uses call-by-name evaluation strategy and reduces $\lambda$-terms to $\beta$-normal form. This machine has three sections in its memory: the term area where the $\lambda$-terms to be performed are written, the stack and the heap. We denote by $\&$t the address of the term t in the term area. In the heap, we have objects of the following kinds:
\begin{itemize}
\item environment: a finite sequence (e, $\xi_{1}$ , . . . , $\xi_{k}$ ) where e is the address of an environment (in the heap), and $\xi_{1}$, . . . , $\xi_{k}$ are closures. There is also an empty environment.
\item closure : an ordered pair ($\&$t, e) built with the address of a term (in the term area) and the address of an environment.
\end{itemize}
The elements of the stack are closures. Intuitively, closures are the values which $\lambda$-calculus variables take.

When term is performed, firstly it is turned to a compiled form. In the compiled form, all bounded variables are replaced with an ordered pair of integers $\langle v, k\rangle$. Where \textbf{v} is a depth of a term and \textbf{k} is a number of an argument. For example, $\lambda$x$\lambda$y(y)x is compiled to $\lambda$2([1, 2])[1, 1], square brackets are used to distinguish from parentheses. $\lambda$2 just shows how many arguments a term has. $\lambda$x($\lambda$y(y)x)x compiled to $\lambda$1($\lambda$1([1, 1])[2, 1])[1, 1], here we can see how the first integer is used in the pair.

\label{section:execution}
The execution then starts with entered a term in the term area (T), empty stack (S) and empty environment (E). Later there are three possible cases in $\lambda$-calculus:
\begin{itemize}
\item Execution of (t)u. We push the closure ($\&$u,E) to the top of the stack and we continue by performing t: thus T points now to t and E does not change.

\item Execution of $\lambda$x$_{1}$...$\lambda$x$_{n}$t where t does not begin with a $\lambda$; thus, T points to $\lambda$x$_{1}$. A new environment $(e, \xi_{1}, \ldots, \xi_{n})$ is created: $e$ is the address of E, $\xi_{1}$, \ldots, $\xi_{n}$ are “popped”: we take the n top entries off the stack. We put in E the address of this new environment in the heap, and we proceed by performing t: thus T points now to t.

\item Execution of x (a $\lambda$-calculus variable). We fetch as follows the value of the variable x in the environment E: in fact, it is a bound occurrence of x in the initial term t$_{0}$. Thus, it was replaced by an ordered pair of integers $\langle v, k\rangle$. If v = 0, the value we need is the k-th closure of the environment E. If v $\geq$ 1, let E$_{1}$ be the environment which has its address in E, E$_{2}$ the one which has its address in E$_{1}$, etc. Then, the value of x is the k-th closure of E$_{v}$ . This value is an ordered pair (T', E') which we put in (T, E).
\end{itemize}

The intuitive meaning of these rules of execution is to consider the symbols $\lambda$x, (, x of $\lambda$-calculus as elementary instructions:
\begin{itemize}
\item "$\lambda$x" is: "pop" in x and increment the instruction pointer.
\item "(" is: "push" the address of the corresponding ")" and increment the instruction pointer.
\item "x" is: go to the address which is contained in x
\end{itemize}

\subsection{Parallelism and distribution}
There are different ways to parallelize and distribute computation. The straightforward approach is to run another machine in parallel when computation is splitted. In this technique we are not using shared memory, but the problem is that as a result we want to run computations on different machines. In order to add this property, we use the actor model. 

The actor model is a model of concurrent computation where \textbf{actors} are universal primitives of concurrent computation. In response to a message that it receives, an actor can: make local decisions, create more actors, send more messages, and determine how to respond to the next message received. Actors may modify their own private state, but can only affect each other through messages (avoiding the need for any locks). This model could be compared to microservices, described in \cite{DragoniLLMMS17}, but instead of services as a unit of system processes (actors) are used.

In the actor system everything is treated as an actor (represented by processes), this actors are is similar objects in object oriented programming. Every actor can receive messages and it can react to these messages in several ways:
\begin{itemize}
\item Send a finite number of messages to other actors
\item Create a finite number of new actors
\item Designate the behaviour to be used for the next message it receives
\end{itemize}
And for these actions no order is assumed, these actions can be performed in parallel. The only sequence is a message queue, which every actor has.

To design the system, different approaches could be used. We can have a fixed amount of actors that will do computations. The main problem of that approach is deadlocks' possibilities. For example, consider if have only two actors and the first one asks the second to compute a term, then the second one can ask the first one to compute a term too, next there would be a situation where each actor waits for the other. Instead, we can create actors on the fly. For this purpose should be introduced a manager actor that lives when an interpreter is launched. This manager actor will create actors to do computations which will die when a computation is finished. In order to make that approach efficient, the language used for implementation should support user space threads, thus actors could be created and killed without large overhead.

\begin{figure}[h!]
  \includegraphics[width=\linewidth]{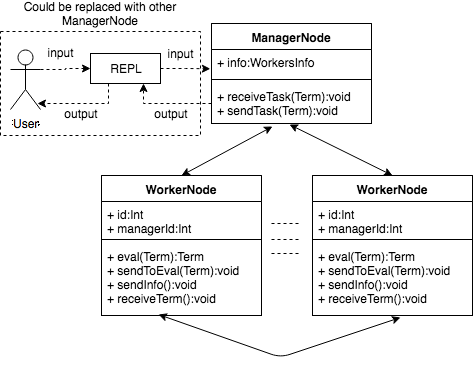}
  \caption{Interpreter scheme in the actor model}
  \label{fig:actors}
\end{figure}

On figure \ref{fig:actors} it is shown how interpreter in the actor model will work. REPL stands for 'read eval print loop' it is an interactive environment where users can enter $lambda$-terms and see how they are computed. There are two types of actors in the system, the first one is a \textbf{WorkerNode} which performs actual computation, and there is a \textbf{ManagerNode} which tracks information about computations (using \textbf{info} table). \textbf{ManagerNode} creates workers to compute terms.
When a user sends a term to compute, a \textbf{ManagerNode} receives the message and creates \textbf{WorkerNode} which should perform this task. The task is sent to a \textbf{WorkedNode} and while it tries to evaluate the term it could decide to split the job (do it in parallel) - in this case the node sends a task to \textbf{ManagerNode}. Manager creates a new worker that will perform the task and send it, when the worker has completed the task he sends back the result to a worker that issued the task. When whole computation is done, \textbf{WorkerNode} sends result back to the manager, then result passed back to the REPL and outputted for a user.

Scaling is done easily, the optimal amount of \textbf{WorkerNodes} is close to the amount of CPU cores available. And if we want a distributed version of this scheme, we can replace \textbf{User} with other \textbf{ManagerNode} which creates other \textbf{WorkerNodes} using different CPU. If we consider computation on one machine, then we even do not need \textbf{ManagerNode}.

\section{Implementation}
\label{chap:impl}
As the Design section was splitted in subsections, this section follows similar structure where each subsection describes implementation of each interpreter component.

\subsection{Read Evaluate Print Loop}
There are several conditions which should satisfy an interactive user environment. Firstly, it should be possible to edit terms entered to REPL, and it is also convenient to have an opportunity to easily repeat previously entered terms by pressing up the arrow key. 

The second problem that should be solved is handling non-terminating terms, consider \textbf{fix f = f fix f} which could be implemented in $\lambda$-calculus as $\lambda$f($\lambda$x(f)(x)x)$\lambda$x(f)(x)x. This term could not terminate. We can solve it by introducing two treads in REPL, the first one handles outputs and the second one handles inputs.

In general, when a user uses a combination of symbols or special keys we should track these actions and produce events internally. Each event could be handled individually. For example, if a user press the left arrow, we should produce event \textbf{MoveLeft} which is handled by moving cursor (a point where user writes text) one position left. When a user press \textbf{Ctrl+C} we should send a message to \textbf{ManagerNode}, explaining that we are not interested in the result of the commutation any more and then \textbf{ManagerNode} must interrupt further computations, by killing actors which computes terms.

\subsection{Parser}
Firstly, a general monadic parser should be implemented according to the Design section. Then it is possible to define a type that represents $\lambda$-terms
\begin{lstlisting}
type Var = String
data Term =
    Variable Var
    | Abstraction Var Term
    | Application Term Term
\end{lstlisting}
This type just naturally repeats $\lambda$-term definition. For each part of this type it is possible to define corresponding parsers \textbf{parseVar}; \textbf{parseAbstraction}; \textbf{parseApplication} and finally \textbf{parseTerm} could be defined using the alternation operator:
\begin{lstlisting}
parseTerm :: Parser Term
parseTerm = parseVar 
            +++ parseAbstraction 
            +++ parseApplication
\end{lstlisting}
Thus, if a parsing result is an empty list, it means that an incorrect term was given.

In order to check that parser works as we expect, we can use several test cases. For parser, firstly should be considered the base cases:
\begin{itemize}
\item x should be \textbf{Variable x}
\item $\lambda$xx should be \textbf{Abstraction x (Variable x)}
\item ($\lambda$xx)z should be \textbf{Application (Abstraction x (Variable x)) (Variable z)}
\end{itemize}
From that point we check a more complicated case $\lambda$f($\lambda$x(f)(x)x)$\lambda$x(f)(x)x that should be \textbf{Abstraction f (Application (Abstraction x (Application (Variable f) (Application (Variable x) (Variable x)))) (Abstraction x (Application (Variable f) (Application (Variable x) (Variable x)))))}

\subsection{Interpreter}
Firstly, before the interpretation, a term should be transformed to compiled form, where bounded variables are replaced with an ordered pair of integers $\langle v, k \rangle$. For free variables we just hold their name. This compiled form enables us not to care about variable names' ambiguity and from the compiled form we can construct $\alpha$-equivalent term to the original one. The compiled form also could be described as a type:
\begin{lstlisting}
data CTerm =
    CVariable Int Int
    | FreeVariable Var
    | CApplication CTerm CTerm
    | CAbstraction Int CTerm
\end{lstlisting}
Compiled form is computed by an induction on the length of term t.
If t = x, we set v = 0. If t = (u)m and the occurrence of x we consider is in u (resp. m), then we compute v, and possibly k, in u (resp. m). Let t = $\lambda$x$_{1}$...$\lambda$x$_{n}$u with $n > 0$, u being a term which does not begin with a $\lambda$: If the occurrence of x we consider is free in t, we compute v in t by computing v in u, then adding 1; if this occurrence of x is bound in u, we compute v and k in u; finally, if this occurrence is free in u and bound in t, then we have x = x$_{i}$, we compute v in u, and we set k = i.

To test the compilation, it is possible to use these test cases:
\begin{itemize}
\item x should be \textbf{Constant x}
\item $\lambda$xx should be \textbf{CAbstraction 1 (CVariable 1 1)}
\item ($\lambda$xx)z should be \textbf{CApplication (CAbstraction 1 (CVariable 1 1)) (Constant z)}
\item $\lambda$f($\lambda$x(f)(x)x)$\lambda$x(f)(x)x should be \textbf{CAbstraction 1 (CApplication (CAbstraction 1 (CApplication (CVariable 2 1) (CApplication (CVariable 1 1) (CVariable 1 1)))) (CAbstraction 1 (CApplication (CVariable 2 1) (CApplication (CVariable 1 1) (CVariable 1 1)))))}
\end{itemize}

After compiling a term it is possible to evaluate it on the Krivine machine where the machine is implemented according to definition. After the evaluation of a term, a compiled form is returned back to readable form, by replacing pair of integers with generated name. The next step is to test the evaluation and it is possible to use the following test cases:
\begin{itemize}
\item x should be \textbf{x}
\item $\lambda$xx should be $\lambda$xx
\item We should test a case where non-termination is possible if the interpreter is not working correctly. Consider ($\lambda$xz)($\lambda$x(x)x)$\lambda$x(x)x, it should just return z.
\item For testing cases which produce intermediate results we can just take a random amount of output and check if it equals to the same amount of repetitions of string. For example, if a random number is 3, we check that by evaluating $\lambda$f($\lambda$x(f)(x)x)$\lambda$x(f)(x)x and taking the 3 first subterms we get (f)(f)(f). 
\end{itemize}

\subsection{Parallelism and distribution}
In the Design section there a general scheme of interpreter was described, but some questions where not discussed. How \textbf{WorkerNode} can issue a task to compute part of the term.

Ideally, parallel computation should be n times faster than sequential where n is the number of cors across the whole system. It is not possible in general, but we can achieve a faster evaluation in subset of cases. Intuitively, there are cases where we cannot perform parallel computation and in this scenario a solution should not be so slower than sequential (consider a time overhead for communication in distributed system). The opposite case is when a term computation can be parallelised. The solution of this problem goes beyond this work.

In our implementation we define a case where parallelism could be done following call-by-name semantic. Consider case (...((v)u$_{1}$)u$_{2}$)...u$_{n}$)u, where v is a variable and u$_{1}$...u$_{n}$, u are terms. We can compute terms in parallel.

To be more specific, consider the term (a)($\lambda$y(z)(y)y)$\lambda$y(z)(y)y. This term never terminates and it gets evaluated as a applied to applications of z: (a)(z)(z)(z)... We can extend this example as ($\lambda$w((a)w)(b)w)($\lambda$y(z)(y)y)$\lambda$y(z)(y)y, so for sequential implementation result will be the same, but in parallel we can compute two terms that never terminate and produce result as: ((a)(z)(z)...)(b)(z)(z)... As a result some parts of a term that never terminates could be computed, consider (((x)t)u)v, where x is a variable, t is a non-terminating term and u, v are terms that terminates, than instead of just sequence of terms produced by t we will additionally see the result of u and v.

When we have a term that terminates and could be parallelised, than we can achieve performance improvement. Ideally if we have a term (...((v)u$_{1}$)u$_{2}$)...u$_{n}$)u where total amount of u is equal to available cores in the system, than computational speed would be bounded to the slowest u.

The next problem is non-terminating terms. In this scheme, if \textbf{WorkerNode} will be computing a term which could not be terminated, it will not return the result. We can solve this problem by saying that \textbf{WorkedNodes} will send not all results, but they will send results as they are computed. So the computation can be treated as a stream, where not finished computations are passed across the system.

Our implementation \href{https://github.com/Bassov/krivine-machine}{implementation}\footnote{https://github.com/Bassov/krivine-machine} does not solve the last problem with non-terminating terms, instead processes tries to compute the term and returns result, so we can see just performance improvement in some cases. Also it runs only on one machine, the problem of distribution will be discussed in evaluation section. To implement parallelism we consider cases described in Krivine machine evaluation:
\begin{itemize}
\item Execution of (t)u - no parallelism
\item Execution of $\lambda$x$_{1}$...$\lambda$x$_{n}$t where t does not begin with a $\lambda$ - no parallelism
\item For execution of x (a $\lambda$-calculus variable) - if x is in the Environment then no parallelism, in other case we take stack and execute each term in separate actor unless there is only one closure in the stack. Then results from actors are collected, combined and outputted.
\end{itemize}
For comparison there is also implementation that uses threads for computation, it uses similar strategy.

\section{Evaluation}
Firstly, parallel version was implemented using actors for distributed environment, thus any local process was used as a remote process. In this case when closure was passed for computation it was firstly serialized to binary, it turns out that the process of serialization could take more time than actual computation. As a result our implementation designed for distribution had lower performance than the sequential implementation. For this reason we used only one machine as it does not require serialized data for passing to local processes.

For benchmarks we used 3 implementations, the first one is sequential, the second one that uses actors and the third one uses threads. For actors implementation we consider only usage on one machine. Time measured as a difference in time between start of computation and the end of computation. The time taken to parse a term is discarded. Benchmarks done using processor with 4 Physical cores - 8 Logical cores. Compiler options: "-threaded -rtsopts -with-rtsopts=-N". Every benchmark run 10 times and than median is chosen because median is robust to outliers.

As Haskell is non-strict language it uses techniques to memorize function evaluation results and it does not evaluate a statement if it is possible. To force evaluation of computed term we can print it out, but for big terms sometimes it is not convenient, thus we just use a function which takes an argument and forces evaluation of that argument. 

First case for benchmarking is a term $(\lambda x\lambda y((a)(x)y)(b)(y)x)\lambda f\lambda z(f)^{m})z)\lambda f\lambda z(f)^{n}z$ where $(f)^{m}$ and $(f)^{n}$ means $(f)...(f)$ $m$ and $n$ times. This term is relatively small but by increasing m and n we can achieve exponential grow of computation complexity and amount of output. This term could be evaluated in two threads.

\begin{table}[h!]
\centering
\caption{Exponential term benchmark}
\label{my-label}
\begin{tabular}{|l|l|l|l|l|}
\hline
m & n & Sequential & Actors     & Threads    \\ \hline
6 & 6 & 0.039103s  & 0.084249s  & 0.025751s  \\ \hline
6 & 7 & 0.153964s  & 0.41152s   & 0.117818s  \\ \hline
6 & 8 & 0.766538s  & 1.993117s  & 0.612723s  \\ \hline
7 & 6 & 0.158522s  & 0.441665s  & 0.970691s  \\ \hline
7 & 7 & 0.673627s  & 1.638231s  & 0.451961s  \\ \hline
7 & 8 & 3.294035s  & 8.95253s   & 3.722431s  \\ \hline
8 & 6 & 0.886852s  & 2.194238s  & 0.800504s  \\ \hline
8 & 7 & 2.944869s  & 8.885262s  & 3.915394s  \\ \hline
8 & 8 & 12.000787s & 18.789062s & 15.015453s \\ \hline
\end{tabular}
\end{table}

As we can see from the table sequential implementation is faster and we know that this term could be computed in parallel. Firstly our implementation computes all parallelized terms, than outputs them, in this case every term produces huge output and by storing this output computation speed becomes slower and memory consumption is also increased. On the other side sequential implementation does not store intermediate result and produces output as it goes.

For the next case we use terms of the form $(\lambda y(x)y)(\lambda xx)*\{n\}z$ - $(\lambda y((((((((x)y)y)y)y)y)y)y)y)(\lambda xx)^{n}z$ where n = 2000000. This terms varies on amount of terms that could be computed in parallel and overall complexity, for sequential implementation complexity should grow linearly, but for parallels interpreters it should be nearly constant. And in order to make computation of this term hard we add a lot of identity functions $(\lambda xx)$.

\begin{table}[h!]
\centering
\caption{Identities benchmark}
\label{my-label}
\begin{tabular}{|l|l|l|l|}
\hline
Amount of y & Sequential & Actors     & Threads   \\ \hline
1           & 5.505613s  & 4.782973s  & 7.680152s \\ \hline
2           & 6.718844s  & 6.727571s  & 8.14726s  \\ \hline
3           & 8.103578s  & 7.321352s  & 7.153319s \\ \hline
4           & 12.434773s & 9.124482s  & 7.079578s \\ \hline
5           & 15.752098s & 8.92889s   & 8.757336s \\ \hline
6           & 16.420424s & 9.363593s  & 8.689104s \\ \hline
7           & 19.532009s & 11.63829s  & 8.497902s \\ \hline
8           & 21.037483s & 10.242402s & 9.81138s  \\ \hline
\end{tabular}
\end{table}

On this table we can see expected result, but there is also small decrease of computation speed for the actors implementation when amount of $y$ is increased.

\section{Discussion}
In order to achieve higher performance on parallel implementation it is necessary to implement output of partial result. For now if we consider term $((v)t)u$ where $t$ is a term computed for 5 milliseconds and $u$ is a term compute in 5 minutes, we still have to wait 5 minutes to see the result. To overcome this problem parallel version should use streams as results of computations.

In general it seems that version which uses threads is faster than version, which uses actors. Also for actors we can use only local processes, because serialization of data takes too much time, thus implementation become worse than sequential in terms of computational speed. So if we want to use actor implementation we should introduce more efficient way to pass closures that should be computed. In case of using interpreter on only one machine it is better to use threads for parallelism.

Additionally parallel implementations could be improved in many ways. If we are trying to increase efficiency by reducing the used space and increasing computation speed, it is possible to improve the Krivine machine with call-by-need semantic \cite{friedman2007improving}\cite{ariola1997call}. Thus, arguments will be evaluated once and then shared. The proof of correctness for each component of the interpreter could be a separate research \cite{jourdan2012validating}\cite{windley1991formal}.


\section{Acknowledgements}
The first author would like to thank Nikita Bogomazov and Maya Stayanova for reviewing this work and making it more readable, parents and friends for supporting him during the study.

\bibliographystyle{IEEEtran}
\bibliography{paper.bib}

\end{document}